\documentclass[sigconf]{acmart}
\AtBeginDocument{%
  }

\setcopyright{acmlicensed}
\copyrightyear{2018}
\acmYear{2018}
\acmDOI{XXXXXXX.XXXXXXX}
\acmConference[Conference acronym 'XX]{Make sure to enter the correct
  conference title from your rights confirmation emai}{June 03--05,
  2018}{Woodstock, NY}
\acmISBN{978-1-4503-XXXX-X/18/06}

\usepackage{multirow}
\usepackage{subcaption}

\begin{document}

\title{DAS: Dual-Aligned Semantic IDs Empowered Industrial Recommender System}

\author{Wencai Ye$^{*}$}
\affiliation{%
  \institution{Kuaishou Technology}
  \city{Beijing} 
  \country{China}
}
\email{yewencai@kuaishou.com}

\author{Mingjie Sun$^{*}$}
\affiliation{%
  \institution{Kuaishou Technology}
  \city{Beijing} 
  \country{China}
}
\email{sunmingjie@kuaishou.com}

\author{Shaoyun Shi}
\affiliation{%
  \institution{Kuaishou Technology}
  \city{Beijing} 
  \country{China}
}
\email{shishaoyun@kuaishou.com}

\author{Peng Wang}
\affiliation{%
  \institution{Kuaishou Technology}
  \city{Beijing} 
  \country{China}
}
\email{wangpeng16@kuaishou.com}

\author{Wenjin Wu$^{\dag}$}
\affiliation{%
  \institution{Kuaishou Technology}
  \city{Beijing} 
  \country{China}
}
\email{wuwenjin@kuaishou.com}

\author{Peng Jiang}
\affiliation{%
  \institution{Kuaishou Technology}
  \city{Beijing} 
  \country{China}
}
\email{jiangpeng@kuaishou.com}

\thanks{$^{*}$ Equal contribution. $^{\dag}$ Corresponding author.}
\renewcommand{\shortauthors}{Ye et al.}
\renewcommand{\arraystretch}{1.15}

\begin{abstract}
%
Semantic IDs are discrete identifiers generated by quantizing the Multi-modal Large Language Models (MLLMs) embeddings, enabling efficient multi-modal content integration in recommendation systems. However, their lack of collaborative signals results in a misalignment with downstream discriminative and generative recommendation objectives. Recent studies have introduced various alignment mechanisms to address this problem, but their two-stage framework design still leads to two main limitations: (1) inevitable information loss during alignment, and (2) inflexibility in applying adaptive alignment strategies, consequently constraining the mutual information maximization during the alignment process.

To address these limitations, we propose a novel and flexible one-stage \textbf{D}ual-\textbf{A}ligned \textbf{S}emantic IDs (\textbf{DAS}) method that simultaneously optimizes quantization and alignment, preserving semantic integrity and alignment quality while avoiding the information loss typically associated with two-stage methods. Meanwhile, DAS achieves more efficient alignment between the semantic IDs and collaborative signals,  with the following two innovative and effective approaches: 
(1) \textbf{Multi-view Constrative Alignment}: To maximize mutual information between semantic IDs and collaborative signals, we first incorporate an ID-based CF debias module, and then design three effective contrastive alignment methods: dual user-to-item (u2i), dual item-to-item/user-to-user (i2i/u2u), and dual co-occurrence item-to-item/user-to-user (i2i/u2u).
(2) \textbf{Dual Learning:} By aligning the dual quantizations of users and ads, the constructed semantic IDs for users and ads achieve stronger alignment.
Finally, we conduct extensive offline experiments and online A/B tests to evaluate DAS's effectiveness, which is now successfully deployed across various advertising scenarios at Kuaishou App, serving over 400 million users daily.
\end{abstract}



\maketitle

\section{Introduction}

Kuaishou is a leading short video and live streaming platform in China, boasting over 400 million active daily users. Advertisement recommendation tasks in Kuaishou can be categorized into two paradigms: (1) traditional discriminative recommendation and (2) generative recommendation. Traditional discriminative Recommendation Systems (RS) typically employ a multi-stage pipeline, comprising retrieval, coarse ranking, fine ranking, and re-ranking, where established techniques like DCN\cite{wang2017deep}, DIN\cite{pi2020search}, and SIM\cite{zhou2018deep} are widely adopted. Recently, the rise of Large Language Models (LLMs) has revolutionized RS design through generative approaches, offering three key advantages: (i) streamlined workflow integration, (ii) improved generalization and stability, and (iii) enabling systematic exploration of scaling laws in RS. This shift has spurred significant academic and industrial innovation, exemplified by works such as TIGER\cite{rajput2023recommender}, HSTU\cite{tanner2015actions}, and LC-Rec\cite{zheng2024adapting}.


In real-world advertising RS, both discriminative and generative models predominantly rely on ID-based features, which often fail to capture the rich multi-modal content (e.g., text, image, video) associated with ads and users. Recent advances in MLLMs have significantly improved multi-modal representation learning ability, unlocking new opportunities for more expressive RS frameworks. \textbf{Semantic IDs (SID)}, derived through vector quantization of multi-modal embeddings from MLLMs, have emerged as a pivotal tool for integrating content information into recommendation scenarios. They are typically applied through two fundamental paradigms: (1) As sparse ID features in traditional discriminative RS models, offering advantages such as learnability and explicit feature crossing with other sparse ID features, or (2) As token IDs in generative RS models, offering the advantages of balancing semantic coherence and discriminative capability, reducing the token space, and improving both training and inference efficiency. 

\begin{figure}
    \centering
    \includegraphics[width=1\linewidth]{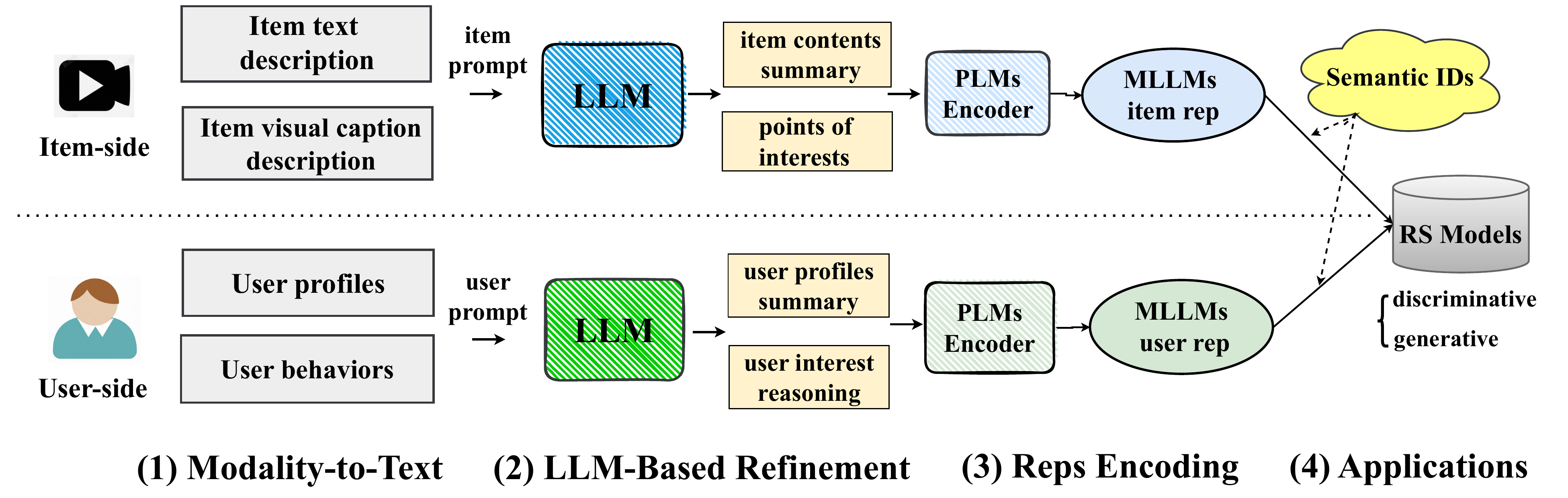}
    \caption{The MLLMs Semantic IDs application pipeline.} 
    \label{fig:mmlms-rep}
    \vspace{-15pt}
\end{figure}

The pipeline for applying SID in advertising recommendation scenarios is illustrated in Fig. ~\ref{fig:mmlms-rep}: (1) Modality-to-Text: Transform raw multi-modal data (user/item profiles, behaviors) into structured textual formats. (2) LLM-Based Refinement: Employ LLMs to generate condensed summaries (e.g., user interest descriptors, item attributes) via their summarization and reasoning capabilities. (3) Representation Encoding: Encode the refined textual content into dense embeddings using Pre-trained Language Models (PLMs). (4) Downstream applications: Utilize these embeddings either as direct dense features or further quantize them into SID for downstream discriminative or generative RS models.

\begin{figure*}[tbh]
    \centering
    \includegraphics[width=1\linewidth]{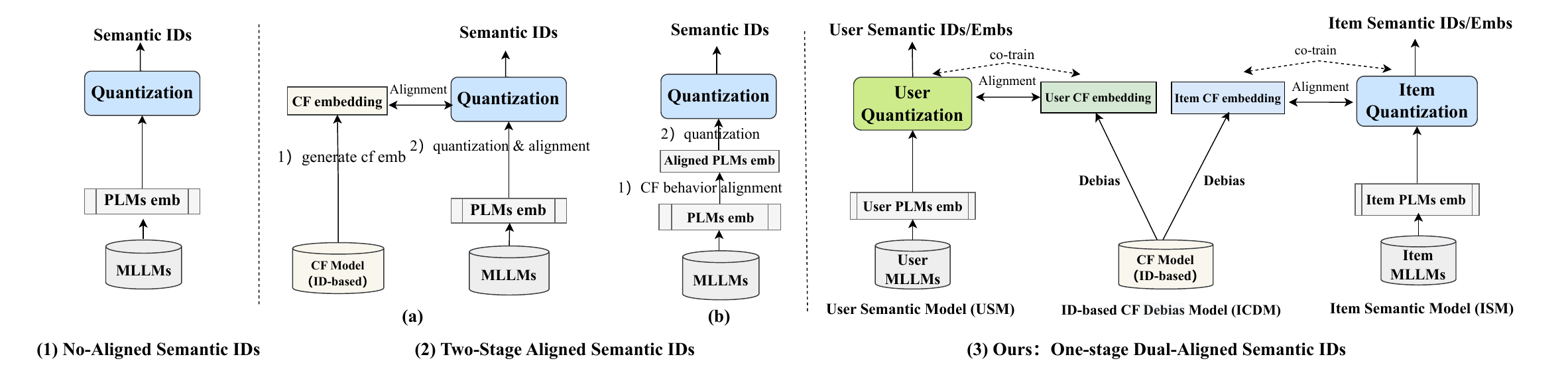}
    \captionsetup{skip=-0.5pt} 
    \caption{Comparison of Semantic IDs construction.
     (1) No-Aligned, (2) Two-Stage Aligned and (3) Ours: One-Stage Dual-Aligned.}
    \label{fig:dual-aligned}
    \vspace{-8pt}
\end{figure*}

The inherent misalignment in consistent objectives between MLLMs content-generated \textbf{No-Aligned Semantic IDs} (as shown in Fig.~\ref{fig:dual-aligned} (1)) and downstream CF-based recommendation tasks has long been a critical challenge in the industry, leading to suboptimal performance when applying SID to downstream recommendation models (e.g.,TIGER\cite{rajput2023recommender} and Zheng et al.\cite{zheng2025enhancing}). 
Therefore, an intuitive yet effective approach is to align the SID with the CF signals to reduce the gap with the downstream recommendation tasks. Recent studies have introduced various alignment mechanisms based on \textbf{Two-Stage Aligned Semantic IDs} (as shown in Fig.~\ref{fig:dual-aligned} (2)) to address this challenge, which can be broadly categorized into two classes:
(a) CF First: A trained CF model generates embeddings, which are then aligned with SID during quantization (e.g., LETTER~\cite{wang2024learnable}). 
(b) Alignment First: Multimodal content representations are aligned using a trained CF model prior to quantization (e.g., QARM~\cite{luo2024qarm}). However, these approaches suffer from three key limitations: (1) The CF and SID models are trained separately, leading to a decoupled optimization process. (2) Quantization and alignment are performed independently, risking information loss and restricting alignment flexibility. (3) Current alignment methods do not maximize the mutual information between the SID and CF signals, ultimately degrading their effectiveness in downstream tasks.

In summary, the main contributions of this paper are as follows:
\begin{itemize}
    \item We present \textbf{D}ual-\textbf{A}ligned \textbf{S}emantic IDs (\textbf{DAS}), a novel one-stage framework that jointly optimizes quantization and alignment through co-training, eliminating information loss in conventional two-stage approaches while supporting diverse quantization and alignment methods.

    \item To achieve more efficient alignment, DAS with the following two innovative and effective approaches:  
    (1) \textbf {Multi-view Constrative Alignment}: To maximize mutual information during CF and SID alignment, we first incorporate an ID-based CF debias module, and then design three effective alignment methods: dual u2i, dual i2i/u2u, and dual co-occurrence i2i/u2u. 
    (2) \textbf{Dual Learning}: By aligning both user-side and ad-side quantizations, DAS generates dual-aligned SIDs with superior alignment.
    \item Extensive experiments validate DAS's effectiveness, showing significant A/B test improvements across all advertising scenarios, with larger gains in cold-start situations. DAS has been fully deployed in Kuaishou's in-app advertising platform, demonstrating strong business values.
\end{itemize}
\begin{figure*}
    \centering
    \includegraphics[width=1.0\linewidth]{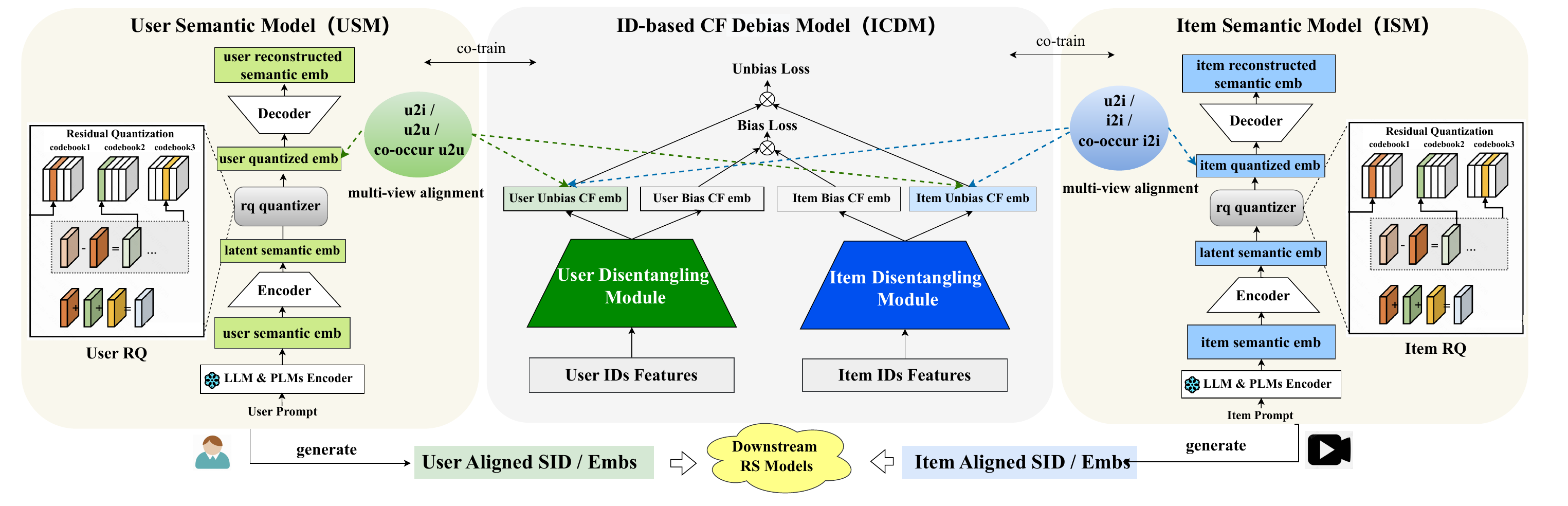}
    \captionsetup{skip=-1pt}
    \caption{The implementation of DAS. UISM module leverages the RQ-VAEs in quantization process, ICDM module uses a disentangled debiasing network to obtain unbiased CF representations, and during the co-training process of UISM and ICDM, alignment between the CF and Semantic IDs is achieved through MDAM module.}
    \label{fig:DAS}
    \vspace{-10pt} 
\end{figure*}
\section{Methodology}


DAS employs one-stage co-training to simultaneously train and align ID-based CF and semantic quantization models, enhancing mutual information between the two, as depicted in Fig.~\ref{fig:dual-aligned} (3). The framework of DAS is also flexible and compatible with various quantization and alignment methods. For example, the quantization component can utilize methods like VQ-VAE\cite{van2017neural}, RQ-VAE\cite{lee2022autoregressive}, or PQ\cite{gray1984vector}, etc., while the CF component can incorporate models such as DSSM\cite{huang2013learning}, GCN\cite{kipf2016semi}, or DCCF\cite{xue2022dccf}, etc. Furthermore, alignment methods act as \textit{plug-and-play} modules to ensure effective alignment between SID and CF representations. 


The specific implementation of DAS is depicted in Fig.~\ref{fig:DAS}, including three core modules: (1) \textbf{U}ser and \textbf{I}tem \textbf{S}emantic \textbf{M}odel (\textbf{UISM}): A quantization-based module that constructs dual SID for users and ads. (2) \textbf{I}D-based \textbf{C}F \textbf{D}ebias \textbf{M}odel (\textbf{ICDM}): Addresses inherent biases in CF signals by learning debiased representations. (3) \textbf{M}ulti-view \textbf{D}ual-\textbf{A}ligned \textbf{M}echanism (\textbf{MDAM}): Performs multi-view contrastive alignment between SID and debiased CF representations for maximizing the mutual information.

\subsection{User and Item Semantic Model}

\subsubsection {Semantic embedding extraction}
To generate accurate user and ad profiles, we leverage LLMs to purify and denoise key information. For user profiles, basic user data and global e-commerce or advertising behaviors are transformed into text to construct a \textbf{user\_prompt}. LLMs are then used for summarization and reasoning to produce refined profile summaries and interest descriptions, denoted as $t_{u}$. Similarly, ad-related data, such as titles, OCR outputs, ASR descriptions, and visual captions, are converted into text to construct an \textbf{item\_prompt}. LLMs generate fine-grained profile summaries, key descriptions, and interest-based selling points, denoted as $t_{i}$. To further utilize these LLM-generated descriptions $t_{u}$ and $t_{i}$ , we employ PLMs for semantic extraction. Specifically, we use the fine-tuned bge m3 model \cite{chen2024bge} to construct MLLMs semantic representations: $s_u\  \epsilon  \ \textbf{R}^{d_1}$, $s_i\  \epsilon  \ \textbf{R}^{d_2}$, where $s_u = \textbf{user\_encoder}(t_u)$ and $s_i = \textbf{item\_encoder}(t_i)$.

\subsubsection {Semantic embedding quantization}
UISM's USM and ISM utilize $s_{u}$ and $s_{i}$ as inputs, respectively, and generate hierarchical SID based on their respective RQ-VAE\cite{lee2022autoregressive}. To prevent codebook collapse in the RQ-VAE, k-means clustering is employed to initialize the codebook. 
The MLLMs semantic embedding $s$ is quantized into the code sequence through the $L$ codebooks, where $L$ is the identifier length. Specifically, for each code level $l\ \epsilon\ \{1,...,L\}$, we have a codebook $C_l = \{e_i\}_{i=1}^N$ , where $e_i\  \epsilon \ \textbf{R}^{d}$ is a learnable code embedding and $N$ denotes the codebook size. Subsequently, the residual quantization can be formulated as:
\begin{equation}
\label{eqn:1} 
\left\{  
     \begin{array}{lr}  
     c_l=\mathop{\text{arg min}}\limits_{i} ||r_{l-1}-e_i||^2,\ \ \ e_i \ \epsilon \ C_l &  \\  
     r_{l} = r_{l-1} -e_{c_l}   
     \end{array}  
     , 
\right.  
\vspace{-2pt} 
\end{equation}
where $c_l$ is the assigned code index from the $l$-th level codebook, $r_{l-1}$ is the semantic residual embedding from the last level, and we set  $r_{0} = s$. Intuitively, at each code level, DAS finds the most similar code embedding with the semantic residual embedding and assigns the corresponding code index as the identifier for the current level. After the recursive quantization, we finally obtain the quantized identifier  $\hat{l}\ \epsilon\ \{c_1,c_2,...,c_L\}$ and the quantized embedding $z$, and then decode it to the reconstructed semantic embedding $\hat{s}$.

The loss for semantic regularization is formulated as: 
\begin{equation}
\small
\label{eqn:2} 
\left\{  
    \begin{array}{lr}  
        \mathcal{L}_{\rm{Sem}} = \mathcal{L}_{\rm{Recon}} + \mathcal{L}_{\rm{RQ-VAE}}, \ \ \ \rm{where},
        \\
        \mathcal{L}_{\rm{Recon}} =||s- \hat{s}||^2,\ \  &  
        \\  
        \mathcal{L}_{\rm{RQ-VAE}} = {\sum\limits_{i=1}^{L}} \  ||sg[r_{l-1}] - e_{c_{l}}||^2   + \mu||r_{l-1}-sg[e_{c_{l}}]||^2,
    \end{array}  
\right. 
\vspace{-4pt}
\end{equation}
where $sg$ is the stop-gradient operation \cite{van2017neural}, and $\mu$ is the coefficient to balance the strength between the optimization of code embeddings and the encoder.

Finally, $\mathcal{L}_{\rm{User\_sem}}$ and $\mathcal{L}_{\rm{Item\_sem}}$ are constructed separately, with the total loss for the SID quantization model represented as:
$\mathcal{L}_{\rm{Sem\_all}}$ = $\mathcal{L}_{\rm{User\_sem}}$ + $\mathcal{L}_{\rm{Item\_sem}}$. 


\subsection{ID-based CF Debias Model}
\begin{figure}
  \begin{minipage}[t]{0.45\linewidth}
    \centering
    \includegraphics[width=1.0\textwidth]{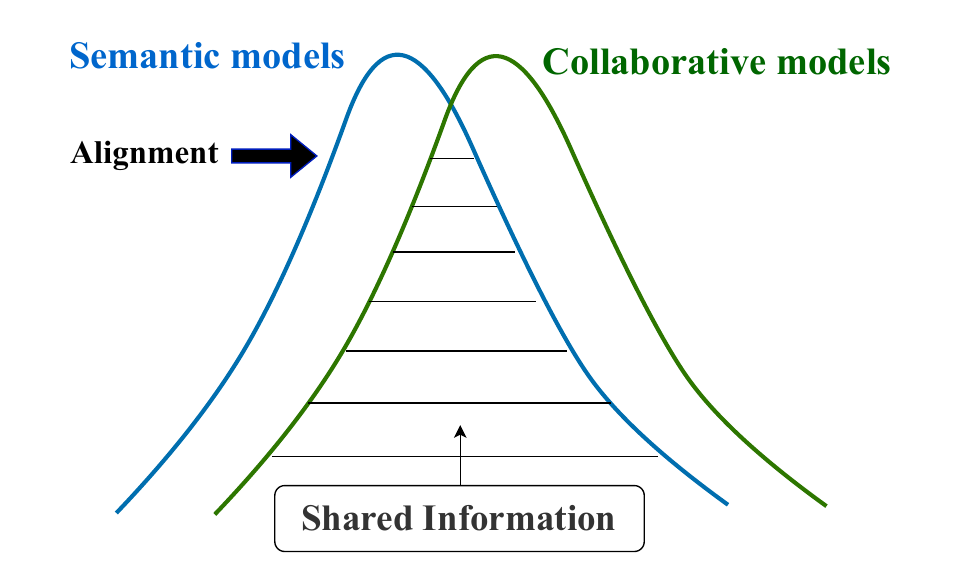}
    \captionsetup{skip=-0.4pt}
    \caption{Semantic and CF \\ models alignment.}
    \label{fig:mutual-info}
    \vspace{-17pt}
  \end{minipage}%
  \begin{minipage}[t]{0.5\linewidth}
    \centering
    \includegraphics[width=1.0\textwidth]{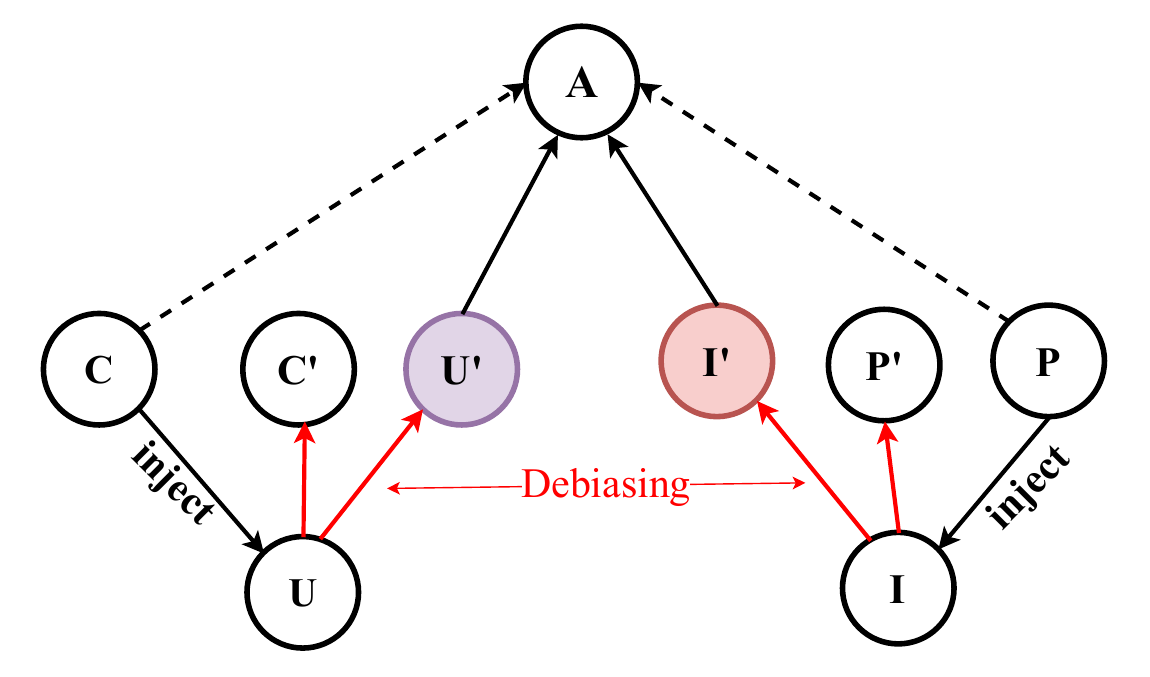}
    \captionsetup{skip=-0.4pt}
    \caption{ID-based CF debias casual graph.}
    \label{fig:casual-graph}
    \vspace{-17pt}
  \end{minipage}
\end{figure}

To incorporate collaborative information into SID and enhance their adaptability to downstream tasks, we align the SID quantization process with the CF model during the co-training. Specifically, we perform cross-view alignment between the SID quantization representations and the ID-based CF representations. The objective is to identify a shared semantic subspace where the two align effectively, as illustrated in Fig.~\ref{fig:mutual-info}. 

In practical advertising settings, ID-based CF models often suffer from severe popularity bias, where 20\% of the ads account for 80\% of the ad revenue, and similarly 20\% of the users contribute 80\% of the ad revenue. On the ad side, this bias manifests itself as popularity bias, while on the user side, it is reflected as conformity bias. When constructing aligned SID based on CF models, it is crucial to prevent the injection of popularity bias into the SID. This bias compromises the integrity of semantic information and hinders the codebook learning process. To address this, an effective approach involves debiasing the CF model first to obtain unbiased CF representations, which are then aligned with the semantic model. Causal inference approaches (e.g., DICE~\cite{zheng2021disentangling}, DCCL~\cite{zhao2023disentangled}, etc.) and disentangling debiasing methods (e.g., \text{$CD^2AN$}~\cite{chen2022co}, etc.) are widely used for debiasing popularity and conformity. 

Therefore, the process of eliminating the ad popularity bias and the user conformity bias can be represented by a causal graph (as shown in Fig. ~\ref{fig:casual-graph}). Let $U$ and $I$ denote the original user and ad representations, and let $A$ be the user’s real action. After the debiasing process, we obtain $U'$ (conformity-debiased user) and $I'$ (popularity-debiased ad), along with disentangled biases $P'$ and $C'$. 

%
Specifically, we adopt a joint training framework~\cite{chen2022co} that leverages a disentangling domain adaptation network to debias both the user and ad representations.
The \textbf{U}ser \textbf{D}isentangling \textbf{M}odule (\textbf{UDM}) separates the representation of user unbiased interest $c_{u}^{int}$ from the user's conformity representation $c_{u}^{con}$ within the user's ID-based features. Similarly, the \textbf{I}tem \textbf{D}isentangling \textbf{M}odule (\textbf{IDM}) separates the representation of the unbiased content of the ad $c_{i}^{pro}$ from the representation of the popularity of the ad $c_{i}^{pop}$ within the ID-based features of the ad. In this setup, $c_{u}^{int}$ and $c_{i}^{pro}$ are encoders composed of MLPs that capture user interest and ad content representations, respectively. Meanwhile, $c_{u}^{con}$ is a conformity encoder composed of MLPs that capture the user's conformity representation, and $c_{i}^{pop}$ is a popularity encoder composed of MLPs that capture the ad's popularity representation. Two constraints guide UDM and IDM optimization toward intended directions:
\captionsetup{font=small}
\begin{equation}
\small
\mathcal{L}_{\rm{user\_item\_sim}}  = {\sum\limits_{b=1}^{B}} \ 1-\frac{c_{u_{b}}^{con}(c_{u_{b}}^{c})^T}{\Vert c_{u_{b}}^{con}\Vert \cdot\Vert c_{u_{b}}^{c})\Vert} +1-\frac{c_{i_{b}}^{pop}(c_{i_{b}}^{p})^T}{\Vert c_{i_{b}}^{pop}\Vert \cdot\Vert c_{i_{b}}^{p})\Vert},
\vspace{-4pt}
\end{equation} 
\begin{equation}
\small
\mathcal{L}_{\rm{user\_item\_orthogonal}} = {\sum\limits_{b=1}^{B}} \ \frac{\Vert c_{u_{b}}^{con}(c_{u_{b}}^{int})^T\Vert ^2}{\Vert c_{u_{b}}^{con}\Vert\cdot \Vert c_{u_{b}}^{int}\Vert} + \ \frac{\Vert c_{i_{b}}^{pop}(c_{i_{b}}^{pro})^T\Vert ^2} {\Vert c_{i_{b}}^{pop}\Vert\cdot \Vert c_{i_{b}}^{pro}\Vert},\\ 
\end{equation}
where $\Vert\cdot \Vert$ represents the norm operation for $L2$ regularization, $c_{u}^{c}$ represents the user's real conformity representation generated by an MLP from user conformity features (e.g., the user's preference for popular ads), and $c_{i}^{p}$ represents the ad's real popularity representation generated by an MLP from item popularity features (e.g., impressions and clicks). The biased user representation $c_{u}$ is defined as $\textbf{MLP}(\text{concat}(c_{u}^{int}, c_{u}^{c}))$, while the biased ad representation $c_{i}$ is defined as $\textbf{MLP}(\text{concat}(c_{i}^{pro}, c_{i}^{p}))$. With $B$ as the batch size, the total disentangling loss is given by: $\mathcal{L}_{\rm{cf\_disentangeled}} = \mathcal{L}_{\rm{user\_item\_sim}} + \mathcal{L}_{\rm{user\_item_\_orthogonal}}$.

The loss for the ID-based CF model consists of two parts: $\mathcal{L}_{\rm{bias}}$ and $\mathcal{L}_{\rm{unbiased}}$, both of which employ batch sampled-softmax loss. Positive samples represent the user's true posterior behavior rate (user's ad click behaviors), while negative samples are obtained through in-batch negative sampling.

The final CF loss is $\mathcal{L}_{\rm{CF\_all}} =  \mathcal{L}_{\rm{cf\_bias}} + \mathcal{L}_{\rm{cf\_unbias}} + \gamma \mathcal{L}_{\rm{cf\_disentangeled}}$, and $\gamma$ is the coefficient to balance the cf and disentangled loss.

\subsection{Multi-view Dual-Aligned Mechanism}
Based on the UISM, we derive SID representations on the user and ad side ($z_{u}$, $z_{i}$) by sum-pooling the codebook embeddings of their hierarchical code indices.
The ICDM generates unbiased user interest and ad content representations ($c_{u}^{int}$, $c_{i}^{pro}$). To bridge ID-based CF and SID representations, the MDAM module introduces three effective multi-view alignment tasks to maximize mutual information between the two.

 

\subsubsection{Dual U2I Constrative Alignment}
Leveraging the observed user-item action relationship  $<U, I, A>$  ($U$: user representation, $I$: ad representation, $A$: user-ad action), this task aims to enhance collaborative information learning for both $z_{u}$ and $z_{i}$ by explicitly modeling and aligning their representations.


We first align the user's quantized SID representation $z_{u}$ with the ad's CF debiased content representation $c_{i}^{pro}$, and this alignment loss $\mathcal{L}_{\rm{a\_u2i\_z_u}}$ is constructed as follows: 
\begin{equation}
\small
-\frac{1}{N}{\sum\limits_{(z_{u},c_{i}^{pro})\epsilon D}^{}}\log\frac{\text {exp}(<z_{u}{^+},c_{i}^{pro}{^+}>)}{\text {exp}(<z_{u}^{^+},c_{i}^{pro}{^+}>) + {\sum_{i=1}^{L}} \text {exp}(<z_{u}{^+},c_{i}^{pro}{^-}>)}.
\vspace{-4pt}
\end{equation}
where $D$ is the set of positive examples within a batch ($A$ is the user's ad click actions) and $L$ random negative examples are sampled within the batch.

Similarly, from the dual perspective, we then align the user's CF debiased interest representation $c_{u}^{int}$ and the ad's quantized SID representation $z_{i}$, which alignment loss $\mathcal{L}_{\rm{a\_u2i\_z_i}}$ is constructed:
\begin{equation}
    \small
    -\frac{1}{N}{\sum\limits_{(c_{u}^{int},z_{i})\epsilon D}^{}}\log\frac{\text {exp}(<c_{u}^{int}{^+},z_{i}{^+}>)}{\text {exp}(<c_{u}^{int}{^+},z_{i}{^+}>) + {\sum_{i=1}^{L}} \text {exp}(<c_{u}^{int}{^+},z_{i}{^-}>)},
\vspace{-4pt}
\end{equation}

\subsubsection{Dual U2U/I2I Constrative Alignment}
To bridge the gap between the quantized SID representations and the CF debiased representations, we perform cross-perspective contrastive alignment, aiming to maximize their mutual information.

 



From the user perspective, we align the quantized SID representation $z_{u}$ with the CF debiased interest representation $c_{u}^{int}$. The alignment loss $\mathcal{L}_{\rm{a\_u2u\_z_u}}$ is formulated as follows:
\begin{equation}
    \small
    - \frac{1}{B}{\sum\limits_{b=1}^{B}} \ \log\frac{\text {exp}(<z_{u_{b}},c_{u_{b}}^{int}>)}{{\sum_{j=1}^{B}} \text {exp}(<z_{u_{b}},c_{u_{j}}^{int}>)}.
\end{equation}

Similarly, from the ad perspective, we align the quantized SID representation $z_{i}$ with the CF debiased content representation $c_{i}^{pro}$, which alignment loss $\mathcal{L}_{\rm{a\_i2i\_z_i}}$ is:

\begin{equation}
    \small
    - \frac{1}{B}{\sum\limits_{b=1}^{B}} \ \log\frac{\text {exp}(<z_{i_{b}},c_{i_{b}}^{pro}>)}{{\sum_{j=1}^{B}} \text {exp}(<z_{i_{b}},c_{i_{j}}^{pro}>)}.
\end{equation}
\subsubsection{Dual Co-occurrence U2U/I2I Constrative Alignment}
Between quantized SID representations, a contrastive alignment is performed based on real co-occurrence relationships. Specifically, the quantized SID representations of ads that co-occur with the same user are encouraged to be closer, and similarly, the representations of users that co-occur with the same ad are aligned.

For two users, $z_{u}$ and $y_{u}$, who interact with the same ad, $\mathcal{L}_{\rm{a\_co\_u2u\_z_u}}$ is constructed as follows:
\begin{equation}
    \small
    -\frac{1}{N}{\sum_{(z_{u},y_{u})\epsilon D}^{}} \log\frac{\text {exp}(<z_{u}{^+},y_{u}{^+}>)}{\text {exp}(<z_{u}{^+},y_{u}{^+}>) + {\sum_{j=1}^{L}} \text {exp}(<z_{u}{^+},y_{u_{j}}{^-}>)},
\end{equation}
where $D$ is the set of positive click examples, and the co-occurrence relationship is recorded based on the memory bank \cite{he2020momentum}.
Similarly, for two ads, $z_{i}$ and $y_{i}$, that are interacted with by the same user, $\mathcal{L}_{\rm{a\_co\_i2i\_z_i}}$ is constructed as follows:
%
\begin{equation}
    \small
    -\frac{1}{N}{\sum_{(z_{i},y_{i})\epsilon D}^{}}\log\frac{\text {exp}(<z_{i}{^+},y_{i}{^+}>)}{\text {exp}(<z_{i}{^+},y_{i}{^+}>) + {\sum_{j=1}^{L}} \text {exp}(<z_{i}{^+},y_{i_{j}}{^-}>)},
\end{equation}
The final multi-view alignment loss is defined as: 
$\mathcal{L}_{\rm{Align\_all}} = 
    \mathcal{L}_{\rm{a\_u2i\_z_u}} + 
    \mathcal{L}_{\rm{a\_u2i\_z_i}} + 
    \mathcal{L}_{\rm{a\_u2u\_z_u}} + 
    \mathcal{L}_{\rm{a\_i2i\_z_i}} + 
    \mathcal{L}_{\rm{a\_co\_u2u\_z_u}} + 
    \mathcal{L}_{\rm{a\_co\_i2i\_z_i}}$.
The total loss of DAS is calculated as follows: 
$\mathcal{L}_{\rm{All}} = 
    \mathcal{L}_{\rm{Sem\_all}} + 
    \alpha\mathcal{L}_{\rm{CF\_all}} + 
    \beta\mathcal{L}_{\rm{Align\_all}}$,
where $\alpha$ and $\beta$ are the hyperparameters controlling the corresponding loss weights.
\subsection{Application of DAS}
\label{sec:app-DAS}
\subsubsection{Feature enhancements for traditional discriminative RS models}
\label{sec:app-DAS-1}

Leveraging the inference capabilities of DAS, dual-aligned SID ($sid_u$, $sid_i$), SID's Embs ($semb_u$, $semb_i$) are generated. Subsequently, we construct the following four types of features, serving as multi-modal content enhancements, which can be seamlessly integrated into traditional discriminative recommendation pipeline models, including the retrieval, coarse ranking, and ranking stages. 

\textbf{(1) ID-based features}: Using the user's $sid_u$ and ad's $sid_i$, learnable prefix-ngram sparse ID features are derived to represent hierarchical SID. For example, if $sid_i$ for ad $i$ is [2,31,142], the features are constructed as ad\_l1=2, ad\_l2=2\_31, and ad\_l3=2\_31\_142, the user-side is similar to the ad-side.

\textbf{(2) List-wise ID-based features}: Based on the ad's $sid_i$ and the user's historical behavior ad sequences, the learnable SID list features are constructed. For example, if the user has clicked [ad1, ad2] with corresponding SID [76, 12, 67] and [76, 22, 242], the list features are: user\_l1\_sids=[76, 76], user\_l2\_sids=[76\_12, 76\_22], user\_l3\_sids=[76\_12\_67, 76\_22\_242].

\textbf{(3) Cross ID-based features}: Learnable sparse ID features are constructed based on the match count of SID between the user's historical behavior sequences of ad's SID and the candidate ad's $sid_i$. For example, if the user history list is [76\_12\_67, 76\_22\_242] and the candidate ad's $sid_i$ is [76, 22, 131], the match count sparse features are: cross\_cnt\_l1=2, cross\_cnt\_l2=1, and cross\_cnt\_l3=0.

\textbf{(4) Dense-based features}: Frozen dense representation features are constructed using the user's $semb_u$ and the ad's $semb_i$.

\subsubsection{Token IDs for generative RS models}
\label{sec:app-DAS-2}
Building on the TIGER\cite{rajput2023recommender} generative task, which uses a multilayer transformer-based encoder-decoder autoregressive structure, we introduce TIGER++, incorporating two key enhancements: (i) User ID is upgraded to $sid_u$; (ii) Ad SID aligned to $sid_i$. For the top-k SID predictions of TIGER++, we have designed two methods for application in the advertising RS: (1) as hierarchical tags for tag-retrieval strategy, (2) as feature enhancements for existing RS models.
\section{System Deployment}
\begin{figure}[tbp]
    \centering
    \includegraphics[width=1\linewidth]{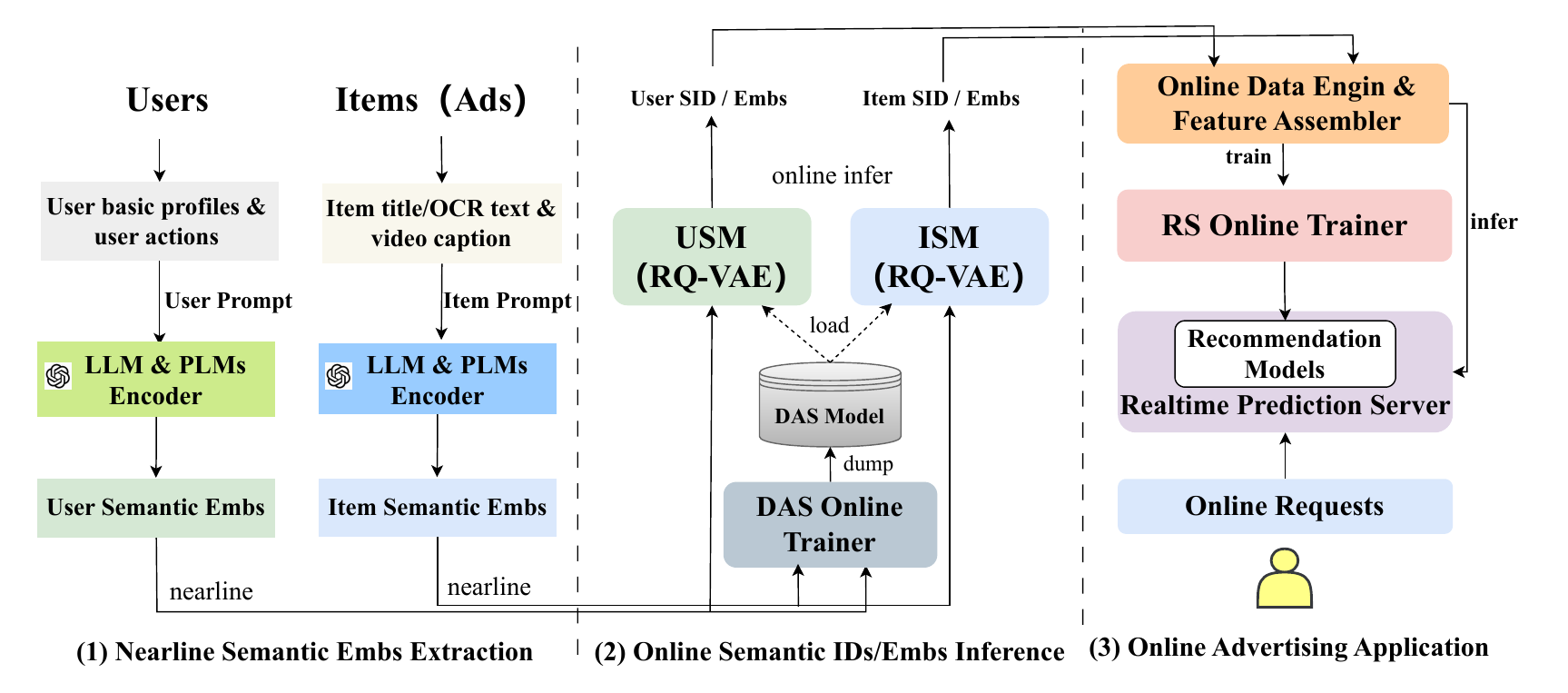}
    \caption{The online deployment pipeline of DAS at Kuaishou.}
    \label{fig:deployment}
    \vspace{-12pt} 
\end{figure}

DAS has been successfully deployed on Kuaishou’s online advertising system, as illustrated in Fig.~\ref{fig:deployment}. 
The deployment pipeline consists of three key stages: 
\textbf{(1) Nearline MLLMs Semantic Embs Extraction}: User profiles and behavioral sequences are first converted to textual prompts, further summarized, and then processed via an LLM to produce refined textual descriptions. The PLMs encoder finally extracts semantic embeddings for users and stores them nearline for the subsequent stage. Similar processing is conducted on the item (ad) side. 
\textbf{(2) Online Aligned SID/Embs Inference}: MLLMs semantic embs are quantized in the DAS online trainer and co-trained with the CF model. In the inference phase, by integrating the USM and ISM modules, it dynamically generates hierarchical SID/SID's Embs in real time, facilitating downstream recommendation tasks. 
\textbf{(3) Online Advertising Application}: The SID/SID's Embs of the user and item are used as auxiliary features for model training and real-time predictions in the recommendation system.
%

\section{Experiment}
We evaluate the effectiveness of DAS through extensive offline and online experiments on real-world industrial-scale data from the Kuaishou advertising platform. The evaluation includes traditional discriminative CTR and generative tasks. 
\begin{table*}[tbh]
\caption{Offline Performance. The "\textbf{Merge Features}" column clearly differentiates between two experimental setups:  "\textbf{Yes}" indicates that user and ad semantic features are merged within a single CTR model, while "\textbf{No}" indicates that they are used in two independent CTR models.}
\vspace{-7pt}
\label{tab:overall}
\begin{tabular}{c|l|c|ccc|ccc}
\hline
\multirow{2}{*}{Alignment Mechanism} & \multicolumn{1}{c|}{\multirow{2}{*}{Model Variants}} & \multicolumn{1}{l|}{\multirow{2}{*}{Merge Features}} & \multicolumn{3}{c|}{Enhanced User-Side Features} & \multicolumn{3}{c}{Enhanced Ad-Side Features} \\ \cline{4-9} 
                                     & \multicolumn{1}{c|}{}   & \multicolumn{1}{l|}{}  & AUC      & UAUC      & GAUC        & AUC      & UAUC     & GAUC     \\ \hline
\multirow{3}{*}{No}                  & CTR Model             & Yes                    & 0.8037   & 0.7736    & 0.7422      & 0.8037   & 0.7736   & 0.7422   \\
                                     & +MLLMs Emb               & No                     & 0.8038   & 0.7738    & 0.7424      & 0.8038   & 0.7739   & 0.7425   \\
                                     & +SID (Raw RQ)           & No                     & 0.8041   & 0.7741    & 0.7426      & 0.8043   & 0.7743   & 0.7428   \\ \cline{1-3}
Two Stage                            & +ASID (LETTER)          & No                     & 0.8044   & 0.7746    & 0.7432      & 0.8046   & 0.7747   & 0.7432   \\ \cline{1-3}
\multirow{4}{*}{One Stage}           & +ASID (Biased DAS)     & No                     & 0.8048   & 0.7749    & 0.7433      & 0.8050   & 0.7748   & 0.7434   \\
                                     & +ASID (Debiased DAS)   & No                     & 0.8050   & 0.7749    & 0.7434      & 0.8052   & 0.7753   & 0.7437   \\
                                     & +ASID (No-dual DAS)    & Yes                    & 0.8054   & 0.7757    & 0.7442      & 0.8054   & 0.7757   & 0.7442   \\
                                     & \textbf{+ASID (Full DAS)} & \textbf{Yes}        & \textbf{0.8061}   & \textbf{0.7764}    & \textbf{0.7448}      & \textbf{0.8061}   & \textbf{0.7764}   & \textbf{0.7448}   \\ \hline
\end{tabular}
\vspace{-7pt}
\end{table*}
\subsection{Evaluation Setup}
\subsubsection{Datasets.}
Public datasets are excluded due to misalignment with the online platform's requirements, limiting their suitability for industrial-scale RS evaluation. Therefore, all our experiments are conducted on billion-scale real-world data sourced from the Kuaishou advertising platform. 
%
%
%
Specifically, for the CTR task, we collect 1.8 billion user-ad interaction samples from approximately 100 million users interacting with 8 million distinct ads. The data sets are split in chronological order with an 80\% training and 20\% testing sets, where positive samples are defined as "ad click".
For the generative task, we use sequences of user-ad interaction of the past 15 days with lengths $\geq$ 10 (filtered to ensure semantic coherence and model efficiency), which yield approximately 50 million users.
We employ a \textit{leave-one-out} strategy \cite{rajput2023recommender} for this dataset splitting.
%


%
%
\subsubsection{Evaluation Metric.}
For CTR tasks, we use AUC, UAUC, and GAUC~\cite{luo2024qarm} for offline evaluation. Additionally, we employ HitRate@K and NDCG@K \cite{rajput2023recommender} as primary metrics to evaluate the performance of generative tasks.
%
%
%
%
\subsubsection{Implementation Details.}
\textbf{(1) Details of the DAS:} The MLLMs semantic embedding dimensions are $d_1$=256 for ads and  $d_2$=1024 for users.
In the RQ-VAE model, the number of codebooks $L$ is 3, each codebook containing $N$=512 codes with a code embedding dimension $d$ of 32. The loss weight coefficients are configured as follows: $\alpha$=1, $\beta$=0.5, and $\gamma$=0.1.
\textbf{(2) Details of the CTR task:} The CTR model integrates SeNet\cite{hu2018squeeze} and MMoE\cite{ma2018modeling}, trained on 16 T4-GPUs using the AdamW\cite{loshchilov2017decoupled} optimizer, with a learning rate of \(1 \times 10^{-4}\) and a batch size of 1024. Each SID sparse feature is represented by a 16-dimensional learnable embedding in the embedding table. 
\textbf{(3) Details of the generative task:} The generative model is developed based on the TIGER architecture, utilizing Google's T5 as the foundational model. 
Specifically, the architecture consists of an encoder-decoder structure, where both components are built from four stacked transformer layers, totaling 13 million parameters. The model was trained on four NVIDIA GPUs over five days to achieve convergence.
%
%
\subsection{Offline Performance}
As shown in Table~\ref{tab:overall}, 
we validate the effectiveness of DAS and its variants by enhancing the CTR model with auxiliary SID features, which solely utilize prefix-ngram ID-based feature settings, as defined in subsection~\ref{sec:app-DAS-1}(1).
%
%

%
The experimental setups are designed as follows: 
\textbf{(1) CTR Model}:
     We employ the online-serving CTR model as the baseline for all experiments, which is an MMoE architecture with SeNet and already incorporates an extensive set of features, including: ID-based features, sequential list-wise features, statistical bucketing features, and frozen multi-modal embeddings.
    %
\textbf{(2) +MLLMs Emb}: 
    Raw MLLMs embeddings are directly used as auxiliary frozen dense features, which are concatenated with other feature embeddings as CTR model inputs.
\textbf{(3) +SID (Raw RQ)}: 
    Refer to \cite{zheng2025enhancing}, SIDs are incorporated into the CTR model as auxiliary prefix-ngram sparse features, which are generated by quantizing MLLMs semantic embeddings via the raw RQ-VAE model.
\textbf{(4) +ASID (LETTER)}: 
    It uses the two-stage alignment model (as shown in Fig. ~\ref{fig:dual-aligned}(a)) to generate aligned SIDs (ASID), similar to LETTER\cite{wang2024learnable}.
\textbf{(5) +ASID (Biased DAS)}: 
    Unlike +ASID (LETTER), we unify the quantization and CF models (without debiasing) into a jointly trained one-stage framework.
    %
\textbf{(6) +ASID (Debiased DAS)}: Replaces CF-biased embeddings of the alignment module with debiased versions through the ICDM module.
\textbf{(7) +ASID (No-dual DAS)}: It integrates SIDs (generated via no-dual quantization models) from both user and ad sides as merged auxiliary features into a unified CTR model.
\textbf{(8) +ASID (Full DAS)}: It uses a unified dual alignment model to simultaneously generate SIDs for the user and ad sides, which are then fed into the CTR model.
Several key insights can be derived from Table ~\ref{tab:overall}:
(i) One-stage methods outperform two-stage methods by jointly training the quantization and alignment models, thereby reducing information loss and enhancing representation consistency.
(ii) +MLLM Emb slightly outperforms the CTR Model, suggesting that frozen MLLM semantic embeddings offer limited benefits due to their unlearnability. Conversely, +SID (Raw RQ) exceeds the CTR Model, highlighting the superior effectiveness of integrating MLLM embeddings into the CTR model in a quantized, sparse, and learnable manner.
(iii) +ASID (Debiased DAS) outperforms +ASID (Biased DAS), validating the ICDM module's debiasing capability for CF embeddings and improving the alignment between semantic and CF representations, reducing popularity/conformity bias influence.
(iv) +ASID (Full DAS) achieves statistically significant improvements over +ASID (No-dual DAS), demonstrating that MDAM's dual-learning framework effectively aligns user and ad SID representations, thereby further improving CTR prediction accuracy.

We also conduct three experiments to evaluate the performance of various SIDs produced by DAS and its variants on the generative TIGER\cite{rajput2023recommender} task, as shown in Table ~\ref{tab:TIGER_task}: 
\textbf{(1) TIGER}: Standard sequential modeling uses raw user ID and no-aligned ad SID sequences (RQ-VAE\cite{lee2022autoregressive}). 
\textbf{(2) TIGER+}: It uses raw user ID and DAS-aligned ad SID sequences. 
\textbf{(3) TIGER++}: Full dual alignment utilizes DAS-aligned user SID and ad SID sequences.
Table ~\ref{tab:TIGER_task} reveals two important findings: (i)  TIGER+ outperforms TIGER, indicating that DAS's alignment mechanism introduces collaborative signals that better link multi-modal content with user behavior, improving the prediction accuracy of next SIDs. (ii)  TIGER++ exceeds TIGER+, suggesting that DAS's hierarchical dual-aligned SID for users provides superior learning benefits compared to conventional user ID.
\begin{figure*}[t]
    \centering
        \begin{minipage}[t]{0.2\linewidth}
            \centering
            \includegraphics[width=0.9\textwidth]{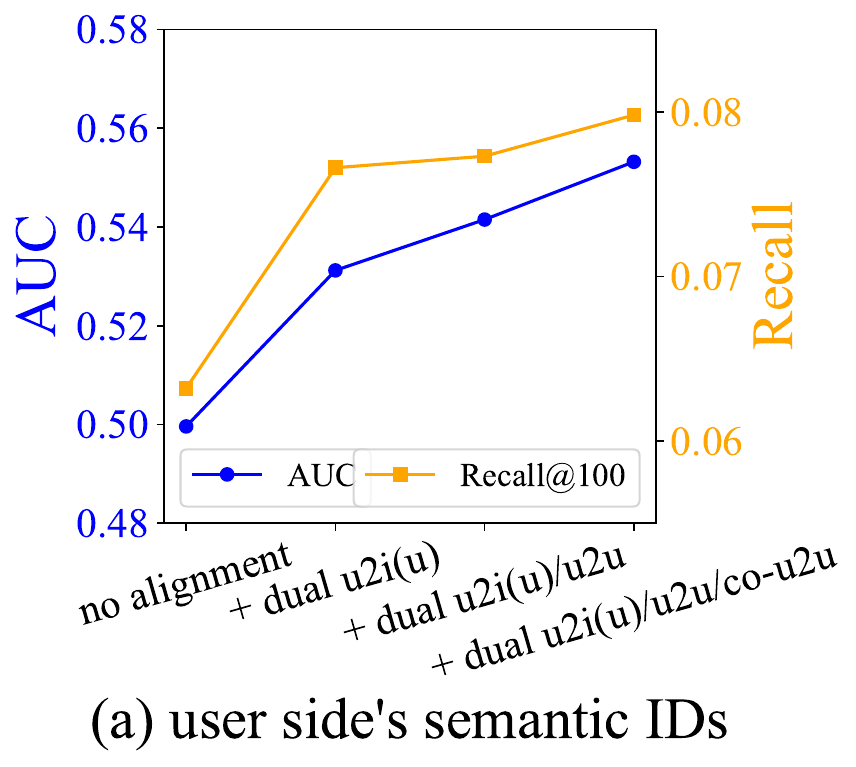}
            \vspace{-13pt}
        \end{minipage}\hfill
        \begin{minipage}[t]{0.2\linewidth}
            \centering
            \includegraphics[width=0.9\textwidth]{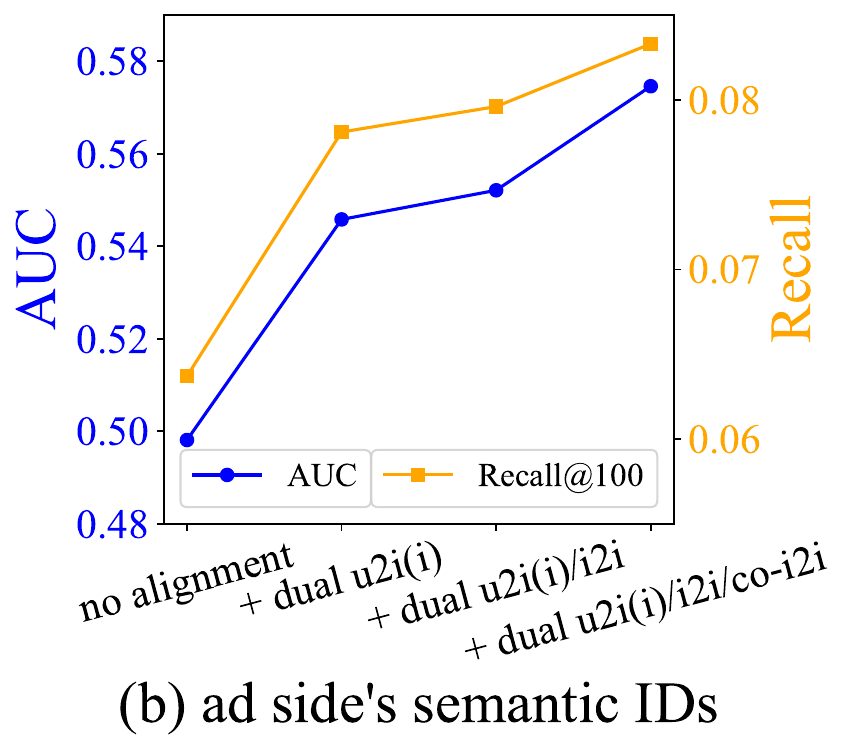}
            \vspace{-13pt}
        \end{minipage}\hfill
        \begin{minipage}[t]{0.3\linewidth} 
            \centering
            \includegraphics[width=0.95\linewidth]{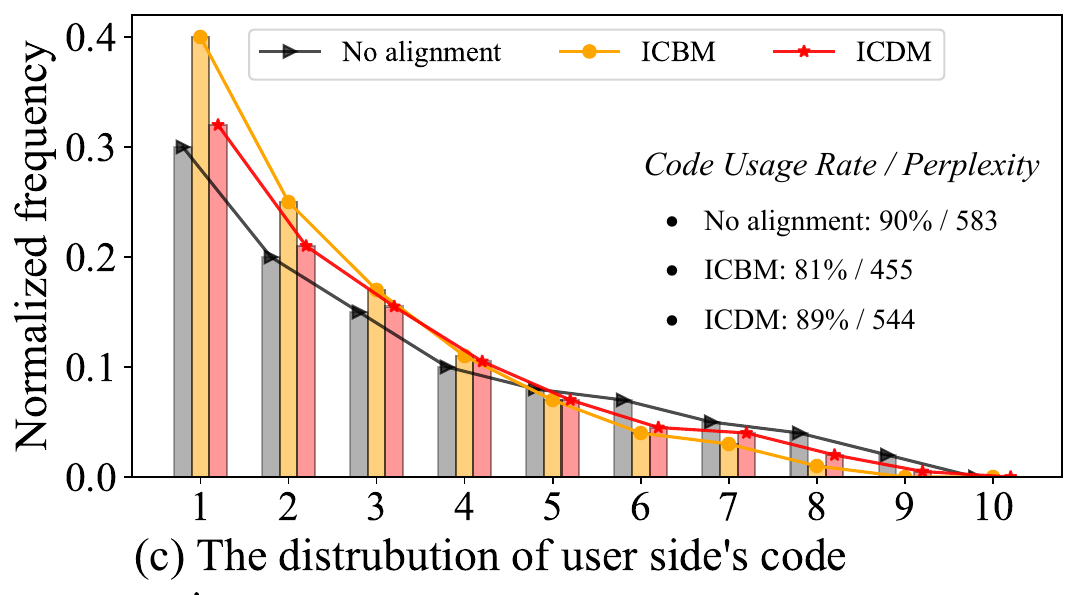} 
            \vspace{-10pt}
        \end{minipage}\hfill
        \begin{minipage}[t]{0.3\linewidth}
            \centering
            \includegraphics[width=0.95\linewidth]{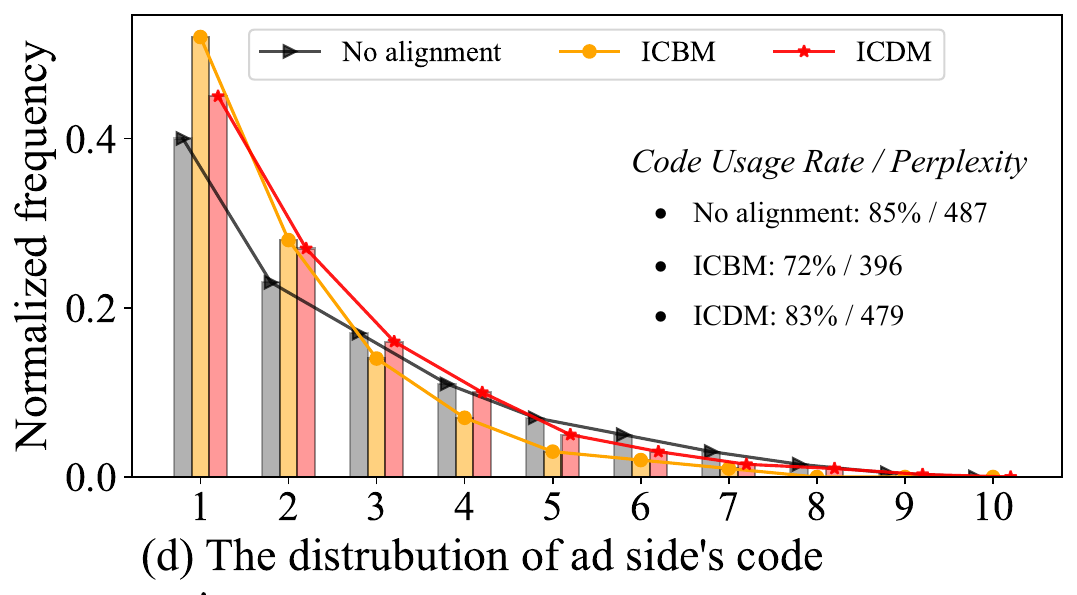}
            \vspace{-10pt}
        \end{minipage} 
    \captionsetup{skip=-0.4pt}
    \caption{In-depth analysis results of DAS. (a) and (b) present the vector retrieval evaluation results for the MDAM module, while (c) and (d) illustrate the code assignment distribution, revealing the performance of the learned codebook. }
    \vspace{-12pt}
    \label{fig:dam-icdm}
\end{figure*}
\begin{table}[th]
\vspace{-10pt}
\caption{Performance of ID Techniques in Generative Retrieval.}
\vspace{-10pt}
\label{tab:TIGER_task}
\begin{tabular}{l|cccc}
\hline
Methods & HR@5        & HR@10       & NG@5     & NG@10    \\ \hline
TIGER   & 0.02184     & 0.02701     & 0.01515  & 0.01712  \\
TIGER+  & 0.02326     & 0.02957     & 0.01611  & 0.01804  \\
\textbf{TIGER++} & \textbf{0.02344}     & \textbf{0.02973}     & \textbf{0.01632}  & \textbf{0.01838}  \\ \hline
\end{tabular}
\vspace{-9pt}
\end{table}
\subsection{In-depth Analysis}
%
\subsubsection{The effectiveness of MDAM}
To verify the effectiveness of MDAM's three alignment mechanisms (dual u2i, dual i2i/u2u, dual co-occurrence i2i/u2u contrastive alignment), we design two evaluation tasks: vector retrieval and downstream CTR prediction.
%

%
To evaluate whether the alignment mechanisms effectively integrate collaborative signals into SID, we design a $\langle u,i \rangle$ vector retrieval task using dot product similarity, evaluated by AUC and recall@100 on users' actual 'ad click' behavior logs.
%
Specifically, we perform the $\langle c_{u}^{int}, z_{i} \rangle$ task to verify the user's collaborative signal integration in ad's SID, where $c_{u}^{int}$ is the user's CF-debiased interest representation and $z_{i}$ is the ad's quantized SID representation. Similarly, the $\langle z_{u}, c_{i}^{pro} \rangle$ task is for the user side, where $z_{u}$ is the user's quantized SID representation and $c_{i}^{pro}$ is the ad's CF-debiased content representation. As shown in Fig.~\ref{fig:dam-icdm}(a) and (b), the progressive stacking of multi-view alignment mechanisms induces successive performance improvements,  highlighting their capability to maximize mutual information between the CF and SID representation.

To further evaluate the alignment mechanisms' capability, we also integrate the user SID and ad SID as auxiliary sparse features into the CTR model. Table~\ref{tab:dam-ctr} shows that DAS consistently outperforms its variants, confirming the effectiveness of the proposed multi-view alignment mechanisms.

\begin{table}[h]
\vspace{-5pt}
\caption{The ablation study of MDAM for CTR tasks}
\vspace{-8pt}
\label{tab:dam-ctr}
\begin{tabular}{l|ccc}
\hline
Model variants                       & AUC       & UAUC     & GAUC     \\ \hline
DAS                                 & 0.8061    & 0.7764   & 0.7448   \\
DAS w.o. dual u2i                   & 0.8055    & 0.7757   & 0.7442   \\
DAS w.o. dual i2i/u2u               & 0.8059    & 0.7761   & 0.7446   \\
DAS w.o. dual co-i2i/u2u            & 0.8057    & 0.7760   & 0.7445   \\ \hline
\end{tabular}
\vspace{-10pt}
\end{table}
%

%
%
\subsubsection{The Impact of ICDM}
\label{subsec:icdm}
To evaluate the impact of CF debiasing in the ICDM module, we design three comparative experiments:
\textbf{(1) No alignment}: Performs quantization using only the raw RQ-VAE without any alignment. 
\textbf{(2) ICBM}: Biased CF representations (\(c_u\) and \(c_i\)) are utilized during the alignment process.
\textbf{(3) ICDM}: Debiased CF representations (\(c_{u}^{int}\) and \(c_{i}^{pro}\)) are used in the alignment process. AS shown in Fig. ~\ref{fig:dam-icdm}(c) and (d), we visualize the distribution of code assignment for the first-level codebook by tokenizing user/ad embeddings with a trained tokenizer, then sorting the first codes by frequency into 10 groups (50 codes/group). Besides, we also calculate the code's usage rate and the perplexity of the first-level codebook \cite{zheng2023online}. 
Finally, we can observe that: (i) incorporating biased collaborative signals reduces code's usage rate and perplexity while simultaneously amplifying the long-tail effect. (ii) ICDM achieves a more uniform code distribution compared to ICBM, effectively mitigating popularity bias and enhancing codebook learning—resulting in superior code utilization rates and higher perplexity scores.
%
%

%
\subsection{Online A/B Test}
%


We conduct a multi-week online A/B test on the Kuaishou advertising platform, randomly assigning 10\% of traffic (approximately 40+ million users), using \textbf{eCPM} (equivalent to revenue) as the primary metric. Rigorous experimentation demonstrates that the proposed DAS method achieves a statistically significant \textbf{3.48\%} improvement in eCPM in all advertising scenarios, with even greater gains of \textbf{8.98\%} observed in cold-start scenarios.

The total revenue benefits from two experimental components: \textbf{For Traditional discriminative RS models}: As summarized in section ~\ref{sec:app-DAS-1}, we design four SID-enhanced features to enhance the performance of the retrieval, coarse ranking, and ranking models in pipelines. 
Together, these features achieve a significant improvement in eCPM \textbf{2.69\%} across various advertising scenarios, highlighting the value of DAS's remarkable practicality in improving online advertising revenue. 
\textbf{For generative RS models}: Top-k SID predictions from TIGER++ to develop two practical deployment strategies for online advertising RS, as summarized in section ~\ref{sec:app-DAS-2}.
This experiment also achieves a remarkable eCPM gain of \textbf{0.79}\%, further validating DAS's effectiveness when integrated into generative recommendation architectures. 
The DAS method currently serves more than 400 million daily active users and is being progressively extended to additional application scenarios.

\section{Related Work}
The recommendation models can be broadly categorized into two major classes: discriminative and generative approaches\cite{li2023large}. In discriminative RS models, multi-modal content is typically incorporated using two main approaches: (1) Direct multi-modal representation fusion: This approach involves aligning and fusing multi-modal representations to construct frozen feature embeddings that can be utilized in downstream models. Representative methods in this category include \cite{sheng2024enhancing,zhang2021mining,chen2024bge,liu2024alignrec}.
(2) Discretization of multi-modal representations: In this approach, multi-modal representations are discretized into semantic IDs using various discretization techniques, and these IDs are treated as learnable features for model input \cite{singh2024better,luo2024qarm,liu2024discrete,zheng2025enhancing}. With rapid advancements in LLM technologies, generative solutions based on the discretization of multi-modal representations into semantic IDs have gained significant traction \cite{rajput2023recommender,tanner2015actions,wang2024learnable,zheng2024adapting,yin2024unleash,jia2025principles}. Among these, quantization methods such as RQ-VAE\cite{lee2022autoregressive} have become widely adopted to discretize multi-modal content into IDs. Other quantization techniques, including those proposed in \cite{van2017neural,esser2021taming,gray1984vector,hou2023learning}, are also extensively explored. To enhance the alignment of multi-modal content with downstream tasks, alignment processes are often required either prior to or during the quantization stage. These alignment techniques aim to optimize the representation for task-specific objectives\cite{ren2024representation,he2022masked,qi2023contrast,yang2024darec}.

%

%
%
\section{Conclusion}
We propose DAS, a one-stage framework that co-optimizes SID generation and CF modeling via joint quantization-alignment training. DAS maximizes SID-CF signal mutual information through: (1) ID-based CF debiasing module and (2) Multi-view dual-aligned strategies (u2i, i2i/u2u, co-occur i2i/u2u). By synchronizing user-ad dual quantizations, DAS enhances alignment precision. Experiments demonstrate significant offline/online gains for both discriminative and generative tasks, deployed at Kuaishou serving 400M+ DAU.


\bibliographystyle{ACM-Reference-Format}
\balance
\bibliography{das_arxiv}

\end{document}